\documentclass[twocolumn]{article}
\usepackage{setspace}
\usepackage{graphicx}
\usepackage[super]{natbib}
\usepackage{amssymb, latexsym}
\usepackage{amsmath} 
\usepackage{amsthm}
\usepackage{bm}
\usepackage[mathscr]{eucal}
\usepackage{enumerate}
\usepackage{multirow} 
\usepackage[center]{caption}

\usepackage{geometry}

\usepackage{authblk}
\usepackage{abstract}

\title{Pitch and timbre discrimination at wave-to-spike transition in the cochlea} 

\author{Rolf Bader \footnote{e-mail: R\_Bader@t-online.de}}

\affil{
	Institute of Systematic Musicology\\ 
	University of Hamburg\\ 
	Neue Rabenstr. 13, 20354 Hamburg, Germany}

\begin{document}
	
	\twocolumn[
	\begin{@twocolumnfalse}
		\maketitle
		\begin{abstract}
			A new definition of musical pitch is proposed. A Finite-Difference Time Domain (FDTM) model of the cochlea is used to calculate spike trains caused by tone complexes and by a recorded classical guitar tone.  All harmonic tone complexes, musical notes, show a narrow-band Interspike Interval (ISI) pattern at the respective fundamental frequency of the tone complex. Still this fundamental frequency is not only present at the bark band holding the respective best frequency of this fundamental frequency, but rather at all bark bands driven by the tone complex partials. This is caused by drop-outs in the  basically regular, periodic spike train in the respective bands. These drop-outs are caused by the energy distribution in the wave form, where time spans of low energy are not able to drive spikes. The presence of the fundamental periodicity in all bark bands can be interpreted as pitch. Contrary to pitch, timbre is represented as a wide distribution of different ISIs over bark bands. The definition of pitch is shown to also works with residue pitches. The spike drop-outs in times of low energy of the wave form also cause undertones, integer multiple subdivisions in periodicity, but in no case overtones can appear. This might explain the musical minor scale, which was proposed to be caused by undertones already in 1880 by Hugo Riemann, still until now without knowledge about any physical realization of such undertones. 
		\end{abstract}
	\end{@twocolumnfalse}
	]

\vspace{2cm}

\section{INTRODUCTION}

 \medskip

Pitch perception is discussed in acoustical terms since Hermann v. Helmholtz had discovered that sounds, and musical tones consist of harmonic overtone series \cite{Helmholtz1863}. Therefore a pitch needed to be caused by the whole spectrum of a sound, not just by a single sinusoidal. Then two tone partials may sound rough when they are close next to each other, because then they cause a beating, an amplitude modulation, where the beating frequency is the frequency difference between the two partials. Helmholtz took 33 Hz to be the frequency difference causing a maximum in roughness perception, and by summing up the calculated roughness of all neighbouring partials he ended in a roughness number of two musical tones played simultaneously. Helmholtz then suggested that the Western music tonal system is based on roughness, where musical intervals are those with least roughness. He found least roughness at simple tone ratios of 2:1 (octave), 3:2 (fifth), 4:3 (fourth) etc.  

Another explanation for the musical major scale is the presence of a major chord in the overtone spectrum of a harmonic tone complex, where the fourth, fifth and sixth partials build a major triad (lower interval is 4:5 as major third, upper interval is 5:6 is minor third). Still there is no minor triad in the overtone series and therefore Hugo Riemann \cite{Riemann1880} proposed an undertone scale as cause of the minor scale, where starting from a fundamental frequency $f_0$, the other partials are integer multiple divisions of the fundamental $f_n = f_0/n$ with n=1,2,3... Then the minor scale appears with n=4, 5, 6 (lower interval is 5:6 as minor third, upper interval is 4:5 as major third). But such an undertone scale was no physically found yet. Still the Trautonium, a musical instrument build by Friedrich Trautwein around 1929 and mainly played by Oskar Sala, uses a subharmonic tone generator which produces untertones.

\medskip

Many studies on pitch perception have been performed during the 50$^{th}$ to the 70$^{th}$ of the 20$^{th}$ century. Maybe most prominently, Schouten \cite{Schouten1962} and later Terhardt \cite{Terhardt1979} proposed a theory of residue and virtual pitch perception at a frequency where a common denominator of the other frequencies of the harmonic spectrum are placed. Still with complex relations between partials, different listeners may perceive residue pitches at very different frequencies. Meeting this finding, Goldstein suggested an optimum processor for pitch perception \cite{Goldstein 1973}. He estimated likelihoods for perceived pitches of two sinusoidals, so that in cases of complex ratios, as e.g. present with inharmonic sounds of percussion instruments, differing pitch judgments of subjects could be explained by a system allowing multiple pitches with respective likelihoods. This model was later enlarged to two tone complexes consisting not only of single sinusoidals but of larger spectra. \cite{Goldstein1978}.

 \medskip

These models discussed pitch in terms of frequency relations. Still when using time series, autocorrelations of these time series were suggested for pitch estimation \cite{Licklider1956}. Similar models were studied with respect to phase, showing phase sensitivity of harmonic and inharmonic sounds and locking of spikes to sound periodicities of stimuli consisting only of a single frequency \cite{Meddis1991}. Such models work very well in terms of Music Information Retrieval for $f_0$ estimation \cite{Klapuri2006}, e.g. with solo singing \cite{Bader2010}. Still here the calculated pitch is not necessarily the perceived one in the light of multiple pitch perception discussed above, as well as in terms of microtonal pitch deviations and intonation, important for instruments where free intonation is possible, like e.g. the singing voice.

 \medskip

With the introduction of neural networks the focus shifted to a spike representation of pitch (see \cite{Cariani2001}, \cite{Lyon1996} for reviews). Here the Interspike Intervals (ISI), the time between two spikes is used for pitch estimation. These theories are referred to as temporal rather than frequency theories. Lyon and Shamma \cite{Lyon1996} propose a temporal theory of pitch perception, also discussing musical intervals of simple integer rations (like 3:2 of the musical fifth or 4:3 of the fourth, etc.). 

 \medskip

For extracting a cochleogram from a time series, gammatone filter banks are most often used (for reviews see \cite{Patterson1995}, \cite{Cariani1999}). Here the transformation from mechanical waves on the BM into spikes uses a Fourier analysis to model the best frequency, spectral filters to model fluid-cilia coupling and hair cell movement, and nonlinear compression to address logarithmic sound pressure level (SPL) modulation. These gammatone filter banks do not use physical modeling of the BM or the lymph and are based on signal processing analogies.

\medskip

Many physical models of the cochlea have been proposed, differing according to the problem addressed in the respective study, like phase-consistency \cite{Steele1979} and frequency dispersion \cite{Ramamoorthy2010}, lymph hydrodynamics \cite{Mammano1992}, active outer hair-cells \cite{Nobili1996} \cite{Szalai2013}, influence of the spiral character of the cochlea \cite{Manoussaki2008}, influence of fluid channel geometry \cite{Parthasarathi2000} among others. The proposed models are 1-D, 2-D, or 3D using the Finite-Difference Method (FDM) \cite{Neely1981}, Wentzel-Kramers-Brillouin (WKB) approximation\cite{Steele1979}, Euler method \cite{Babbs2011}, Finite-Element Method (FEM) \cite{Kolston1996}, Boundary-Element Methods (BEM) \cite{Parthasarathi2000} or a quasilinear method \cite{Kanis1996}. All these models are stationary. Time-dependent models have been proposed to model otoacoustic emissions \cite{Verhulst2010} or to compare frequency and time domain solutions \cite{Kanis1996}. Still all these physical models do not incorporate the transfer from mechanical to electrical energy, and therefore result in estimations of best frequency locations, phase distributions on the BM or enhancements of best frequencies, but not in cochleograms.

 \medskip

The present paper proposes a pitch model based on the spike patterns the mechanical sound wave excites on the BM. It suggests that pitch is the presence of spike periodicities over many bark bands, while timbre is represented by a mix of periodicities distributed over different bark bands. This is in contrast to previous models in two ways. First, the fundamental periodicity is shown to be present at all bark bands with spectral energy, previously only single bands or their sum were discussed. Secondly, such a spike pattern does not need further signal processing calculations, like autocorrelations, the pitch is already there at the cochlea nervous output, and this pitch is clearly distinguished from the other periodicities, which form the timbre. As until now no neural network has been found displaying an autocorrelation, which would need to be present in the nucleus cochlearis or the trapezoid body, pitch perception based on an autocorrelation is of doubt.

 \medskip

Therefore a Finite-Difference Time Domain (FDTD) model of the cochlea with a transfer of mechanical energy into spikes is used as introduced before \cite{Bader2015}. The cochleogram calculated is then further investigated in terms of Inerspike Interval (ISI) histograms and bark Periodicities (BP) to show the temporal and spatial coding of pitch and residue pitch. 

\medskip

To distinguish temporal and spatial representation in the cochlea, according to literature, in this paper the place-theory means the association of places on the BM with respective best frequencies, frequencies which cause the BM to vibrate with a maximum amplitude at these places. Contrary, the temporal or frequency representation is the temporal periodicity of spikes as displayed by the ISI distribution. The use of the term tonotopical representation is omitted here, as in the literature it may be used for both cases or mixtures of them. So e.g. the tonotopical gradient in the primary auditory cortex (A1) (\cite{Moore2001} among many others) sorts best frequencies in terms of their place in the A1, but this term may also be used to denote temporal ISI, especially in the cochlea.
 
\section{METHOD}

\subsection{FINITE-DIFFERENCE TIME DOMAIN (FDTD) MODEL}

The present model has been discussed in detail before \cite{Bader2015}. A phase synchronization of spikes appeared at the transition between mechanical energy on the BM and the spiks excited by this energy. In a harmonic sound the time points of the spikes caused by the lower sound partials locked to the time points of the spikes caused by higher partials. This synchronization appears because of the nonlinear transition condition between BM movement and spike excitation. This transition is well studied (see e.g. \cite{Hubbard1996}). When the mechanical wave on the BM is shearing the stereocilia of a hair cell, a small protein thread, the ip link, between two cilia is stretched and therefore opens a tiny channel in the stereocilia cell membrane where ions flow in and cause a depolarization of the cell. The time point of channel opening is therefore caused by the mechanical wave on the BM reaching a maximum shearing and is implemented in the model using two conditions. A spike at one point X on the BM at one time point $\tau$  is excited if

\newpage

\begin{enumerate}
\item $u(X-1,\tau) < u(X,\tau) > u(X+1,\tau))$, \\
\item$u(X,\tau-1) < u(X,\tau) > u(X,\tau-1))$ \ .
\end{enumerate}

Condition (1) means a maximum shearing of two nervous fibers, as a necessary condition to an opening of an ion channel. This only happens with a positive slope as only then the stereocilia are driven away from each other. With a negative slope the cilia are getting closer and therefore no stress appears at the tip link between them. This corresponds to the rectification process in gammatone filter banks.

 \medskip 
 
Condition (2) is a temporal maximum positive peak of the BM displacement. It is the temporal equivalent to the spatial condition, a maximum acceleration where the tip link between the cell and its neighbouring cells is most active.

 \medskip

A Finite-Difference Time Domain (FDTD) model was implemented as a cochlea model. The basic Finite-Difference model was successfully implemented with musical instruments and \cite{Bader2013} \cite{Bader2005} and is highly stable and reliable. Finite-Difference methods have preciously been used with cochlea models but only as eigenvalue problems \cite{Neely1981}, next to a mother models like Finite-Difference, Euler methods or the like, as discussed above. Still with time-dependent problems FDTD methods are suitable because of stability and realistic frequency representation.

 \medskip

The model assumes the BM to be a rod rather than a membrane, as here it is assumed to be 3.5 cm long and only 1 mm wide at the staple and 1.2 at the apex. So a 1-D model is enough when not taking the spiral or the fluid channels into consideration \cite{Bader2015} \cite{Babbs2011}. The fluid dynamics is neglected because the speed of sound in the lymph, which is about 1500 m/s, is much larger than on the BM which is about 100 m/s at the staple and decreases fast to about 20 m/s at the apex \cite{Ramamoorthy2010}. Therefore for a single sinusoidal the same phase is present all along the BM at one time point, and the long-wave approximation \cite{deBoer1991} can be used.

 \medskip

Then the 1-D differential equation of the model is linear but inhomogeneous with changing stiffness. In the model values of \cite{Allen1977} are used with stiffness $K = 2 \times 10^9 e^{-3.4 x} dyn / cm^3$ changing over length x of the BM. The linear density $\mu$  is only slightly changing along the length because of the slight widening of the BM like $\mu(x) = m / A(x)$. According to the Allen values the mass is assumed constant over the BM like  $m = 0.05 g / cm^2$.

 \medskip

The damping of the BM is also depending on space. According to the time integration method of the FDTD model a velocity decay at each time integration is applied. Using a sample frequency of $s= 200 kHz$ the Allen damping values $d = 199 - .002857 x \ \frac{dyn\ s}{cm^3}$ are implemented using a velocity damping factor of  $\delta(x) = .995 - 0.00142857 x$.

\subsection{REPRESENTATION OF SPIKES}

Four representations are used in the follow.

\subsubsection{WAVE FORM} 

The wave form, as present in the lymph of the cochlea driving the BM. The speed of sound in the lymph is about that of water, which is $\sim$ 1500 m/s, the speed of sound on the BM is much lower, and about 100 m/s at the staple and only about 20 m/s at the apex. Therefore a long-wave approximation \cite{deBoer1991} is implemented in the model which assumes instantaneous impact of the wave form on the whole BM. This is reasonable for all spikes up to 4 kHz, the maximum temporal firing rate of spikes, already using interlocking of several spikes firing together at different time points. There might be a slight impact of the travel time of the wave in the lymph above 10 kHz, still there no pitch perception is present and therefore this impact is neglected here.

\subsubsection{COCHLEOGRAM} 

The spikes caused by the traveling wave over the BM are represented in a cochleogram C. C is a vector, where each spike i is a vector entry $C_i$. Each spike has three parameters, its time point $C_i^t$, its bark band $C_i^B$ and its amplitude $C_i^A$. 24 bands are used. If a spike is caused on the BM, the amplitude of the wave on the BM location is represented as gray-scale in the cochleogram. The FDTD model uses approximately two nodal points for each bark band, depending on the width of the band. Therefore it is expected that the many neurons within a band will contribute to a ensemble burst which is larger with stronger amplitude of the traveling wave on the BM. This relationship is not linear, and therefore the gray-scale representation does not necessarily represent the physiological output strength of the spikes perfectly. Still this strength is not needed in the reasoning of this paper and so the gray-scale presentation is kept as additional information of the amplitude on the BM causing the respective spikes.

\subsubsection{ISI HISTOGRAM} 

The Interspike Interval (ISI) histogram is calculated by accumulating the interspike intervals over all 24 Bank bands and over a time interval T. The accumulated ISI over the time interval $\Delta T$ starting at t of all adjacent spikes i summing over all bark bands B is

\begin{equation}
ISI^f_t = \sum_i  \frac{(C^A_i + C^A_{i+1})/2}{C^t_i - C^t_{i+1}}
\end{equation}

for $t <  C^t_i < t + \Delta T$ summing over all bark bands B. The term $C^t_i - C^t_{i+1}$ is in the denominator and so inverted. Therefore the interspike time intervals are converted into frequencies for convenience. Also the ISI is amplitude weighted as discussed above, here an average amplitude is used. Finally, the ISI is taken over all bark bands B.

As the sample frequency of the cochlea model is 192 kHz, accumulating frequencies with such a high precision is pointless, as it is very unlikely that two interspike time intervals will meet exactly to be possibly accumulated. Therefore an accumulation interval is chosen with width corresponding to the bark band sizes. So the ISI is integrating over all bark bands, but the resulting ISI has discrete integration interval center frequencies f.

\subsubsection{BARK PERIODIGRAM (BP)} 

Another important measure is the presence of spike periodicities and their respective frequencies in each single bark band. Although each bark band has its critical frequency, still spikes need not to be sent out regularly from this band and drop-outs or irregular spiking may happen. Therefore a measure is needed to show the ISI present in each band. This is done here by integrating all $C^t_i$ present over the time of the whole sound but for each band separately between adjacent spikes, but also between spikes further apart like

\begin{equation}
BP^{f, R}_B = \sum_i \sum_{r=1}^{R} \frac{(C^A_i + C^A_{i+r})/2}{C^t_i - C^t_{i+r}}
\end{equation}  

Here r is the separation between two spikes in the integral, where r=1 is the adjacent spike and $r > 1$ is a spike further away. For $r>1$ therefore the BP is summing over all r up to R.

So as the ISI does integrate over all bark bands taking a closer look at the time development, BP integrates over the whole time and displays the frequencies at each bark band. So with the latter the relation between spatial and temporal representation is displayed. This is an interesting measure as at a place on the BM with a certain best frequency, a periodicity may be present with another frequency than this best frequency.

 \medskip

First only intervals of adjacent spikes are calculated. Still also periodicities are present between spikes placed further away. In this case the suggested BP is a sum over multiple neighbours starting from the adjacent spikes with r=1 and ending with a maximum distance R. Below R=1 (only adjacent spikes) and R=10 (up to the tenth neighbour) are used.
 
\subsection{SOUNDS USED IN THE SIMULATION}

Five sounds have been used to investigate the burst-like impulse train, the residue behaviour and the representation of fundamental and partials in the ISI histogram.

\begin{itemize}
\item{TC400} A tone complex of a harmonic overtone series with 10 partials and a fundamental frequency of 400 Hz is used, where the amplitudes of the partials decay with 1/n with partial numbers n=1,2,3,...10.
\item{TC600} The same complex as TC400 is used, still now with a fundamental frequency of 600 Hz.
\item{TTC400600} A two-tone complex consisting of both, TC400 and TC600 as a sum.
\item{TTS400600} a two-tone sound now only consisting of the fundamental partials of TC400 and TC600, so of only two sinusoidals, 400 Hz and 600 Hz.
\item{Guitar} A classical guitar tone with fundamental pitch of 204 Hz recorded with a microphone.
\end{itemize}

\section{RESULTS}

\subsection{WAVE FORM VS.IMPULSE TRAIN}

A spike on the BM can only appear when a wave is traveling over the respective position on the BM at the respective time point. So only when there is sound pressure of the sound time series arriving at the ear, transferred to the BM, there will be a BM displacement, and the sound energy is causing spikes. So with impulse-like time series, bursts of spikes at the time points of the sharp amplitude impulses are expected. Also, as the wave on the BM is traveling from high best frequencies regions on the BM at the staple to low best frequency regions at the apex, an exponential time decay need to be present for such bursts, as is found experimentally \cite{Cariani1999}.

 \medskip

Fig. 1 to Fig. 2 show such bursts of spike patterns caused by the impulses of the time series. Each strong positive slope of the wave form, shown in the wave form plot at the top of each figure, causes spikes all over the BM with the expected exponential decay over time which can be seen in the cochleograms in the middle of the figures. After this impulse, additional spikes appear, but fade out over time, which can clearly be seen in the higher harmonics. So in the spike patterns, the impulse-like time series of the wave form is preserved, and the spikes leave the cochlea in a burst-like manner. 

 \medskip

The simulations are all shown with the initial transient of the BM when starting at rest, and being driven by the time series shown in the wave forms. It appears that in all cases the cochleogram shows a regular periodicity right after the second spike of the fundamental. Therefore the periodicity of a complex tone is represented as a spike interval after only one period of the sound. Higher partials are even represented earlier, as regular periodicities appear with higher partials even within the first periodicity of the sound. Still as the spike representation is very sparse compared to the time series, for a ISI periodicity to appear at the fundamental frequency one whole period is needed.

\begin{figure}
\centering
\includegraphics[width=0.8\linewidth]{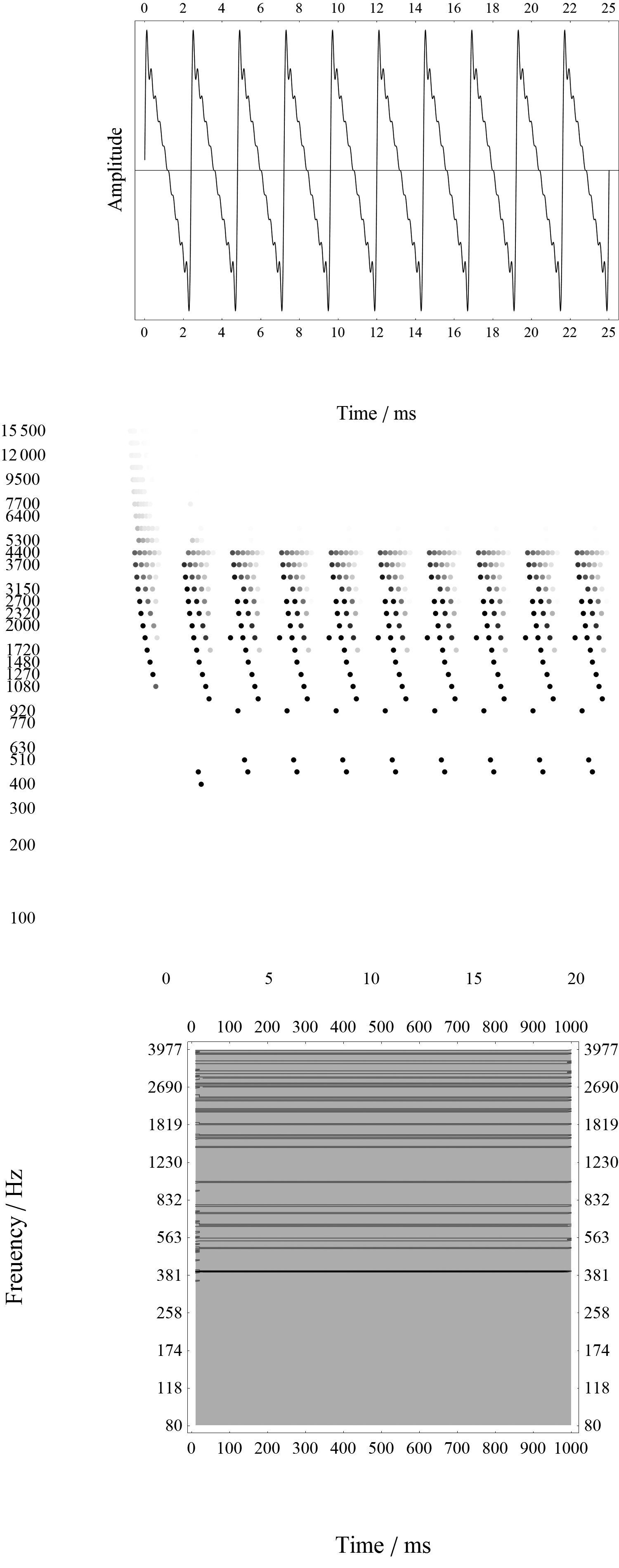}
\caption{Time series (top), cochleogram (middle) and interspike interval (ISI) histogram (botton) of a harmonic tone with fundamental frequency of 400 Hz, consisting of 10 partials with 1/n decaying amplitudes with n=1,2,3,... The cochleogram shows the spike pattern following the amplitude pulses of the time series. The cochleogram shows energy between 400 Hz - 4kHz as expected. The ISI histogram clearly shows 400 Hz as a basic periodicity, still not all higher harmonics are represented, as the ISI of the spikes of a frequency band need not to match the frequency of the respective bands.}
\label{fig:ResidualPitch_400Hz_1_f_Decay_All}
\end{figure}
   
 \smallskip

\begin{figure}
\centering
\includegraphics[width=0.8\linewidth]{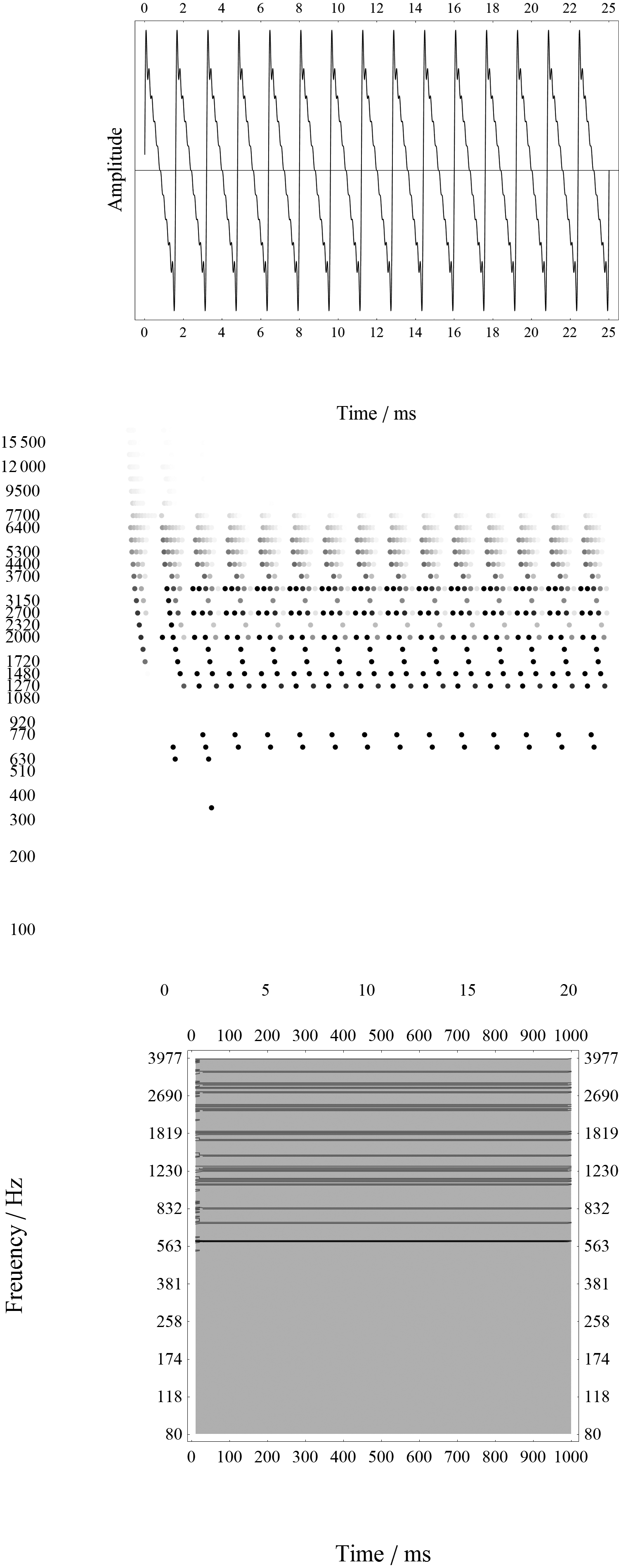}
\caption{Same as Fig. 1 only now for a 600 Hz tone. Again in the cochleogram the energy is between 600 Hz - 6kHz as expected, still the ISI histogram does display several periodicities not matching the higher partials.}
\label{fig:ResidualPitch_600Hz_1_f_Decay_All}
\end{figure}

 \smallskip

\begin{figure}
\centering
\includegraphics[width=0.8\linewidth]{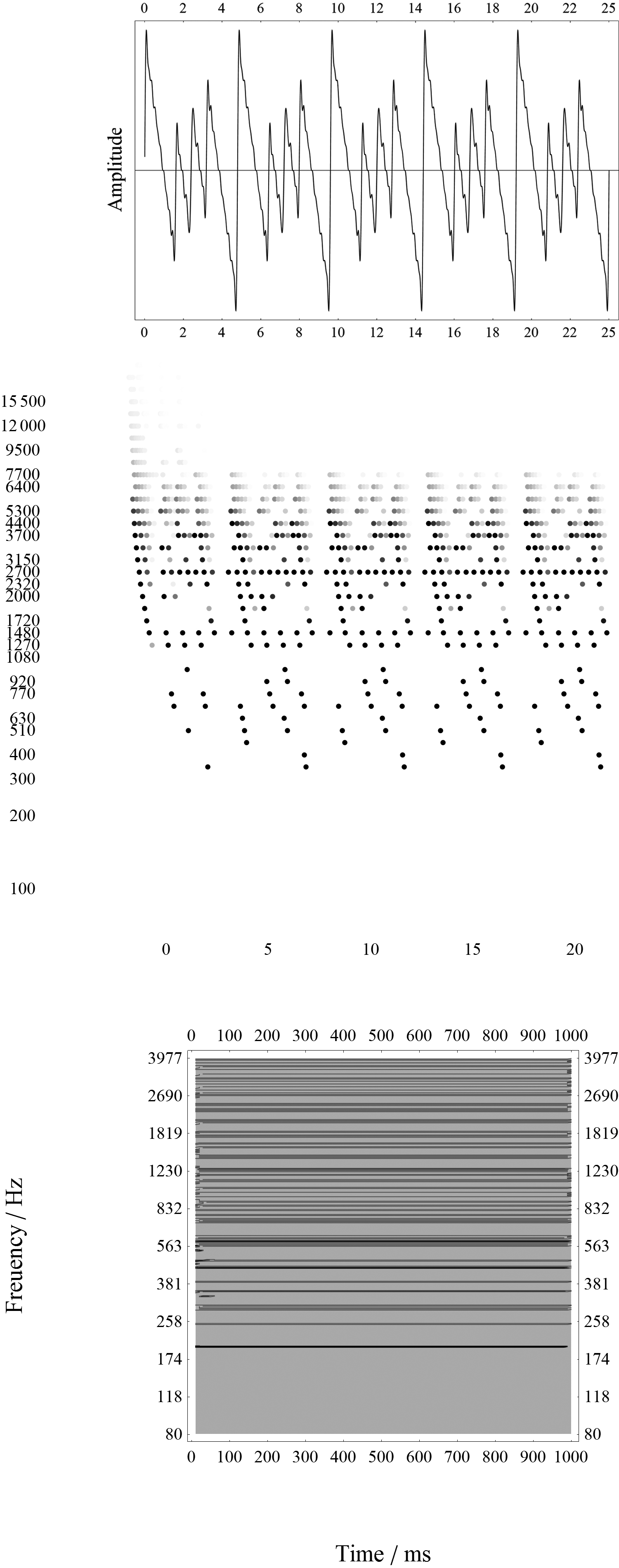}
\caption{Similar than Fig. 1 and Fig. 2 still now with the two tones combined, leading to a musical fifth (3:2). Again the cochleogram spike patterns follow the impulse train of the time series. But now the ISI histogram displays a 200 Hz periodicity as the fundamental frequency, although in the cochleogram there is no energy within the 200 Hz best frequency band. Therefore the expected residue of 200 Hz appears as spike pattern in the ISI histogram but not as energy on the best frequency band on the basilar membrane.}
\label{fig:ResidualPitch_400Hz_600Hz_1_f_Decay_All}
\end{figure}     
       
 \smallskip

\begin{figure}
\centering
\includegraphics[width=0.8\linewidth]{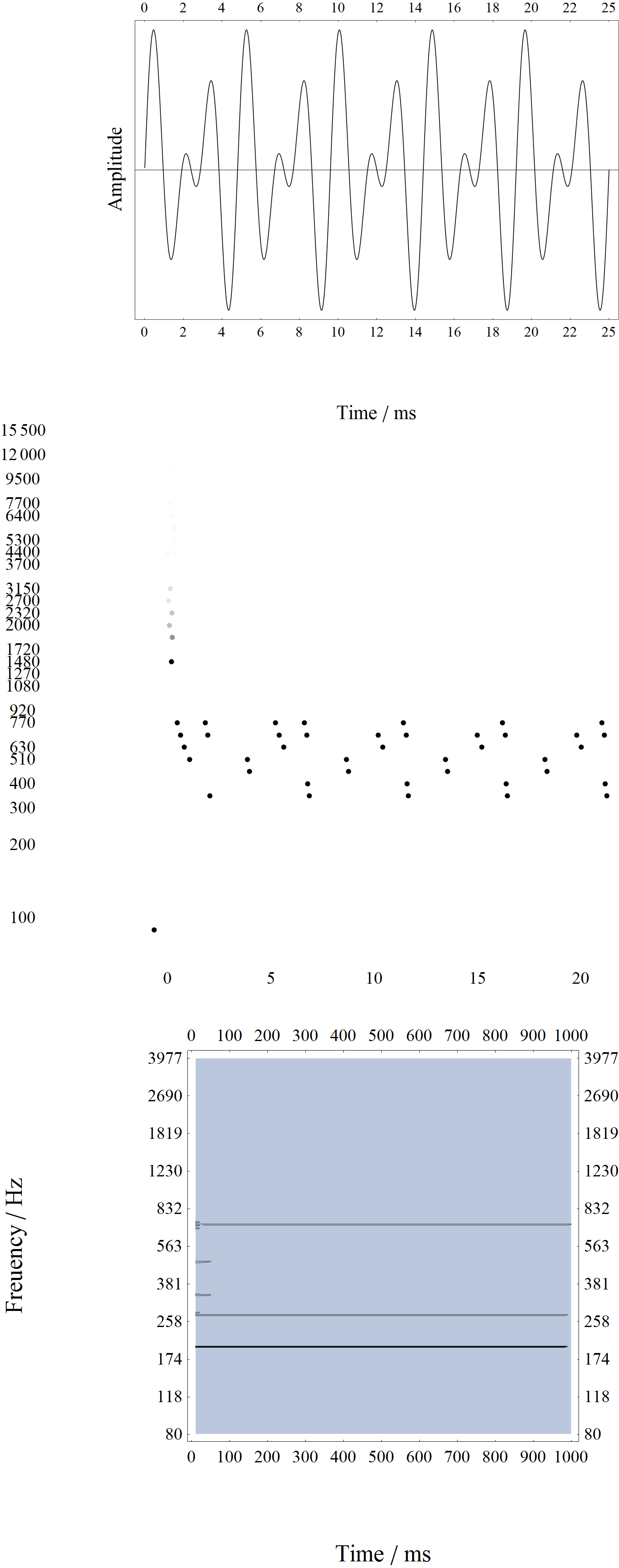}
\caption{Same as Fig. 3 but now for two sinusoidal tones of 400 Hz and 600 Hz. Again in the cochleogram there is no energy at the expected residue frequency of 200 Hz but in the ISI histogram the strongest periodicity is at the 200 Hz residue frequency.}
\label{fig:ResidualPitch_400Hz_600Hz_Sine_All}
\end{figure}

\subsection{RESIDUAL PITCH}

\medskip

Residual pitch is musically relevant in many cases, e.g. for organ builders. Low pitched organ pipes are very long, where e.g. a $\lambda/4$ tube of a A0 pitch of 27.5 Hz need to be 3.13 m. Covering a whole octave with pipes of similar length is expensive. Therefore it is easier to use the residue pitch of two pipes instead, so e.g. producing the A0 pitch from two pipes of E0 and A0. Both tones are the second and third partial of A0, and therefore a A0 pitch is perceived. Another example of a residue pitch appears with church bells. The strike note of these bells is a residue pitch which is heard, but physically not present as a partial.

In the ISI histograms of Fig. 1 to 4 the lowest periodicity present in the sound is represented as a frequency present over time. TC 400 in Fig. 1 shows a lowest frequency at 400 Hz, TC600 in Fig. 2 shows a lowest frequency of 600 Hz. Still TTC400600 in Fig. 3 shows a lowest frequency of 200 Hz. This is the residue frequency of two tones of 400 Hz = 2 $\times$ 200 Hz and 600 Hz = 3 $\times$ 200 Hz. In the cochleogram of TTC400600 in the middle of Fig. 3 such periodicities can clearly be seen as temporal distances between two spikes in one bark band, e.g. in the 400 Hz or between th 920 Hz and 1080 Hz band. Still in the cochleogram there are no spikes in the 200 Hz band.

 \medskip

The same situation appears in the TTS400600 ISI histogram at the bottom of Fig. 4, where the lowest periodicity, and indeed the strongest one is 200 Hz. Again in the respective cochleogram in the middle of the figures no spikes are present at the 200 Hz bark band.

 \medskip

Still now with the TTS400600 case there are no longer periodicities at 400 Hz or 600 Hz, the frequencies the sound is build of. Still the energy of the sound is still present in the 400 Hz and 600 Hz bark bands, so these frequencies are represented spatially although no longer temporally.

 \medskip

Also the ISI histograms only takes adjacent spikes and their periodicities into consideration. When taking several spikes in one Band into consideration, regular patterns or microrhythms may appear which is beyond the scope of this paper.

 \medskip

So a residue pitch is present in the ISI periodicity right at the spike pattern leaving the cochlea in a temporal representation, but it is not present spatially in the respective bark band on the BM.

\subsection{PERIODICITIES IN BARK BANDS}

The ISI historgram discussed above is integrating over all bark bands to show all periodicities and their respective frequencies present during one time interval. Still it is interesting to see which periodicities appear at the different bark bands. Such bark Periodicity (BP) representations as discussed above are shown in Fig. 5. Three cases are displayed, TTS400600 (top), TTC400600 (middle) and a single tone played on a classical guitar with a fundamental frequency of 204 Hz used to meet the residue pitch of 200 Hz found with the TTC400600 and TTS400600.

 \medskip

In all cases the 200 Hz periodicity is widely distributed over many bark bands. The TTS400600 has not much energy, as only two sinusoidals enter the cochlea, still the 200 Hz periodicity is by far the strongest and distributed from the 300 Hz to 720 Hz bands. Although we would expect the energy to be in the 400 Hz and 600 Hz bands, an enlargement of the bands is expected as the spatial representation is often not very sharp.

 \medskip

In the TTC400600 case again the 200 Hz periodicity is distributed over several bands. Still now there are more periodicities as the sounds now have 10 partials each. These additional periodicities are all present only with lower periodicities compared to the best frequencies of the bands. So a bark band may have periodicities with lower frequencies than its best frequency, but do never have higher ones. The TTC400600 case of Fig. 5 (middle) shows spikes still very well present along a line of linear positive slope, where the lowest periodicity of the bark band can respond in correlation to the input sound.

 \medskip

The pitch definition also works with a guitar tone shown at the bottom in Fig. 5. The fundamental frequency of the tone at 204 Hz is present at all bark bands, still with a maximum intensity at the 200 bark band as expected. There is also energy around 91 Hz, also distributed over nearly all bark bands. This frequency is the Helmholtz resonance of this classical guitar, which is present in the sound too. Both pitches are close to each other, still they are not perfectly harmonic. Therefore they do not combine to a residue pitch but appear distinguished one from another.

 \medskip

It is interesting to see in the plot that the fundamental pitch at 204 Hz and the Helmholtz at 91 Hz, as two inharmonic tones are both represented over a large amount of bark bands, while at the same time the higher harmonics of the guitar tone do not appear in such a way. The next two harmonics can be seen at the 400 Hz bark band but with different periodicities, analog to the findings of the TTS400600 case. Therefore the 91 Hz single sinusoidal is another pitch present throughout the sound, and is therefore represented as a second periodicity. This is musically and perceptually correct. The 91 Hz Helmholtz is much lower in volume and therefore the harmonic tone dominates perception. Still the Helmholtz resonance is also audible with normal guitar playing in the low range. The BP is able to find this additional pitch present in the sound.

 \medskip

The BP of the TTC400600 case, shown in the middle of Fig. 5 is similar to that of the guitar. It has a periodicity present at many bark bands at its residue frequency. The other distributed periodicities represent the timbre. They are not perfectly regular due to spike drop-outs, as discussed above. So in these bark bands doubling, tripling, etc. of the periodicity of this band appears. The lines with constant slops in the BP of TTC400600 and the guitar tone are expected and represent the tone partials. As the BP plots are log-log plots, with frequency plotted logarithmically on both axis, higher harmonics will appear as lines with constant slopes.

 \medskip

In summery, the fundamental pitch is represented by one periodicity appearing over many bark bands, while timbre is a distribution of many periodicities over the bands. So when discussing a musical tone as consisting of a pitch and a timbre, this split is justified in terms of the spike pattern leaving the cochlea (now of course not discussing further processing steps in following neural nuclei). Still this split is present at the very first processing stage, and therefore no further processing need to be present. The dominance of pitch perception over timbre perception, as found in many Multidimensional Scaling perception tests \cite{Bader2013}, could be explained by the presence of one periodicity at many bark bands simultaneously, right at the lowest neural processing level.

 \medskip

\begin{figure}
\centering
\includegraphics[width=0.7\linewidth]{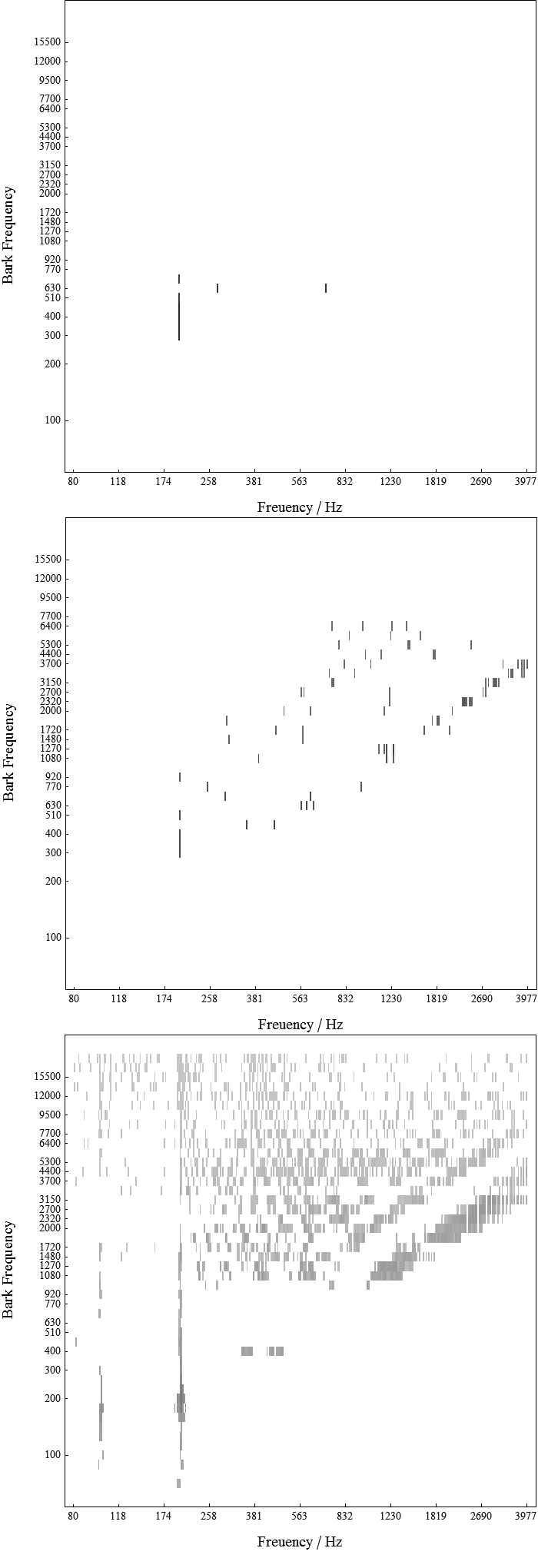}
\caption{Distribution of ISI periodicity over bark bands with sounds of Fig. 4 (top), Fig. 3 (middle) and a classical guitar tone with 204 Hz fundamental (bottom). The pitch ISI periodicity appears over many bank bands, while the partials appear as lines with constant slope in the log-log plot. Although ISI periodicities in bark bands with lower frequency than the bark bands may appear, in no case a bark band fires with a spike frequency higher than the barks critical frequency.}
\label{fig:ResidualPitch_Periodicity_vs_bark}
\end{figure}
 
\medskip

\subsection{MULTIPLE DELAYED bark PERIODICITIES}

The BP discussed above only took the interspike intervals of adjacent spikes into consideration (R=1). To estimate a frequency or periodicity spectrum over longer time intervals the BP for R=10 was calculated. Remember that this means the summery of all possible combinations from R=1 to R=10. Fig. 6 shows BP for R=10 for TC400, TTC400600 and the guitar tone already discussed above with fundamental of 204 Hz. 

 \medskip

The TC400 plot on the top of the Fig. 6 shows the basic principle. The 400 Hz pitch appears again over all bark bands where spectral energy is spatially present. Of course in a multiple delayed BP plot, this fundamental need to repeat in the undertones of 200 Hz, 100 Hz and 50 Hz which is clearly the case. Also again the timbre is represented as periodicities, again only appearing at the respective bark bands and no longer distributed over many bark bands, as is the case for pitch. Reading the plot horizontally, the periodicities of the higher partials appear again as undertones, that is as subdivisions $f_B/n$ with bark frequency $f_B$  and n=1,2,3,... This results in regular patterns between the vertical lines, representing periodicities which are symmetric between two periodicities below the fundamental of 400 Hz, with a logarithmic distortion of spreading out towards higher frequencies. Above 400 Hz this pattern reappears, but is now distributed up to 4kHz, stretched according to the logarithmic scale. Still as expected there is no higher pitch periodicity above 400 Hz (like 800Hz, 1200Hz, etc.) and the pattern stretches out up to 4kHz.

 \medskip

As TC400 is a tone complex of 10 partials with amplitudes 1/n, with partial number n=1,2,3,...,10, there are ISI at the higher partials, which are not distributed over many bark bands. An exception is the second partial n=2 with 800 Hz. Here the symmetric pattern has its mirror axis, and this periodicity is present at several bark bands. Still this does not hold for higher partials $n>2$.

 \medskip

So again the figure shows a fundamental difference between pitch and timbre, where pitch is represented by ISI spatially distributed over all bark bands containing spectral energy, and timbre is represented spatially as a complex pattern, distributed over best frequency BM positions.

 \medskip

Examining the TTC400600 case in the middle of Fig. 6, again the 200 Hz residue pitch is present in nearly all bark bands with spectral energy. Still the two fundamental pitches of 400 Hz and 600 Hz both show alignment of ISI over many bands, although blurred around these frequencies. This also holds for 1200 Hz to some extend. Taking the pattern in higher bark bands into consideration, similar patterns as present in the TC400 (the plot above) can be examined, most clearly maybe with 600 Hz. So these slightly aligned ISI over several bark bands are again the symmetry axis of these patterns, and are present at the fundamentals of the two tone complexes. Therefore they are in between pitch, with a very precise ISI over many bark bands, and timbre, where the ISIs of the partial only appears at the respective spatial place on the BM.

 \medskip

The BP of the guitar shown at the bottom of Fig. 6 shows a very similar behaviour like the TC400 and TTC400600. The guitar pitch of 204 Hz is the only one present over all bark bands. The same holds for the Helmholtz resonance peak of 92 Hz. Around the 204 pitch the expected symmetric pattern appears with the pitch as symmetry axis, the same holds for the 91 Hz pitch. The partials are again present with all their undertones. Additional symmetry patterns align with the partials 400 Hz, 600 Hz and 800 Hz quite clear, still again much more blurred than the pitch ISIs at 204 Hz and 91 Hz.

\medskip

\begin{figure}
\centering
\includegraphics[width=0.7\linewidth]{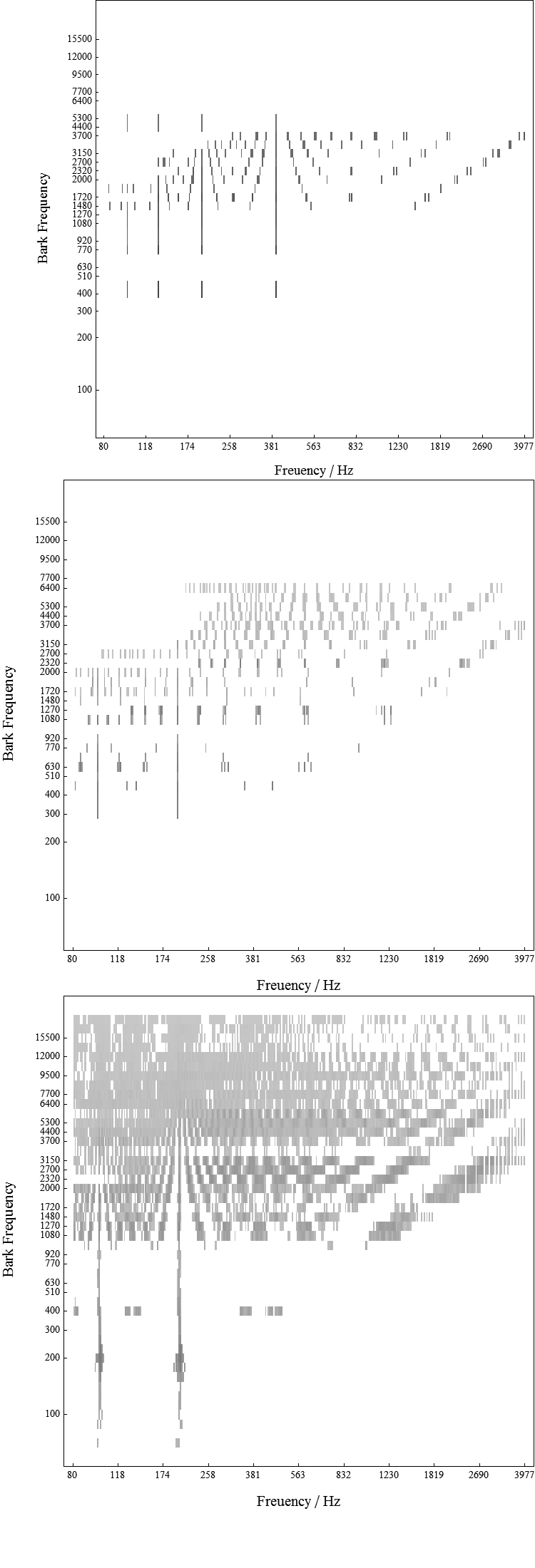}
\caption{BPs as frequencies appearing in bark bands of the ISI histogram, here as a sum of ISI between spike n and n+r with r=1,2,3,...10, for a 400 Hz tone complex from Fig. 2 (top), a 400 Hz + 600 Hz tone complexes from Fig. 3 (middle) and the guitar tone of 204 Hz fundamental from Fig. 5 (bottom). Clearly the pitch appears as periodicity (or frequency) at all bark band where spectral energy is in the sound, while the higher partials only appear in certain bark bands only. Again the difference in coding between pitch and timbre appears also with multiple ISI intervals.}
\label{fig:ResidualPitch_barkPeriodicities_10Delays}
\end{figure}

\medskip

\section{CONCLUSION}

The spike patterns of the cochleogram in all cases clearly have an onset at the largest peak of the wave form, the largest impulse and therefore the spikes are then getting more sparser during one fundamental periodicity of the sound. This is a physical necessity, as a spike can only be excited when the BM has displacement, which again can only be the case when the incoming wave form has a large amplitude. So although a frequency representation of the spike pattern is of interest, the temporal development of the wave form determines the spike pattern. So the spike bursts leaving the cochlea have a maximum at the largest wave form amplitude.

 \medskip

So according to the wave form, and of course according to the energy in the higher sound partials, drop-outs in the regular spiking pattern of the respective partials need to occur. As can be seen in the ISI plots above, especially in the higher partials, such drop-outs do indeed happen, making the representation sparse. These drop-outs produce an undertone spectrum of the respective best frequency, like $f_n = f_0 / n$ with n=1,2,3,... Indeed in no case a best frequency bark band showed frequencies above its best frequency but only at its best frequency or at lower integer sub-frequencies. Therefore the spatial representation of best frequencies at BM locations holds generally, still at each bark band a full or a partly undertone spectrum can also occur. So the temporal and spatial representations are not perfectly separated one from another. 

\medskip

In this context it is interesting to note that such an undertone spectrum has been suggested as a reason for the minor tone scale, analog to the presence of the major scale present in the overtone spectrum \cite{Riemann1880}. Still until now no physical reality of such an undertone spectrum was known. But due to drop-outs of spikes such an undertone spectrum is physically present already at the transition between mechanics and spikes in the cochlea.

 \medskip

This mechanism is also responsible for the fundamentally different representations of pitch and timbre. Pitch, as suggested here, is an ISI periodicity which occurs at nearly all bark bands driven by any present partial in a sound. This periodicity is very narrow banded. Contrary, the periodicities of higher partials, the timbre, are mainly only present at the bark band of their respective best frequency, their places on the BM, along with possible undertone periodicities. Also the timbre ISI periodicities are not that narrow in bandwidth compared to the fundamental pitch. Therefore timbre is considerably differently coded than pitch. With the guitar sound having a single sinusoidal frequency at 91 Hz, inharmonic to the guitar spectrum with fundamental of 204 Hz, this 91 Hz periodicity again is coded over very many bark bands, again with a very narrow bandwidth. Although this frequency is low in the guitar sound, it still is coded the same way as the basic guitar pitch, pointing again to a different treatment of pitch and timbre.

 \medskip

The coding process found does also hold for residue pitch. In the cases of two harmonic spectra with a simple common residue pitch of 200 Hz, this periodicity is clearly visible in the wave form, and the ISI clearly shows the narrow band periodicity at 200 Hz as a residue pitch. Also the BP shows the residue as the pitch in the present sense of a narrowband periodicity over many bark bands. The fundamental pitches of the two tones the sound is constructed of are coded in a wider band, but clearly appear in the BP plot. Therefore a residue pitch already appears temporally, although not at all spatially, as in the 200 Hz bark band no spikes are found. Here a relation between the salience of spatial versus temporal representation for perception might be found, comparing the strength of residue perception with the strength of the ISI coding of the residue.

 \medskip

Examining the initial transients of the cochleograms for the sounds, which are all without transients themselves (except for the guitar sound), the speed of the cochlea to estimate the periodicity is indeed astonishing, where the system takes only one to two periodicities to arrive at a pitch estimation, when the ISI is taken into account.

\section{DISCUSSION}

The findings suggest a definition of pitch as a narrow band ISI periodicity, which is present at nearly all bark bands with energy in the driving sound. Timbre then are ISI periodicities appearing distributed over many bark bands. Therefore, models of coincidence detection in higher neural nuclei might no longer be necessary to understand pitch perception, but this cannot decided here. Also the suggested pitch definition need to be tested with inharmonic sounds and large tone complexes, like musical clusters.

 \medskip

Also synchronization will need to be discussed with such a system. As shown in a previous study\cite{Bader2015} with the present model, synchronization of spike phases on the cochlea appear during the transition from mechanical to electrical spike excitation. Such synchronization also takes place in the nucleus cochlearis and the trapezoid body as shown experimentally \cite{Joris1994} \cite{Louage2004}. This synchronization of spike patterns is necessary to adjust the time delay of spikes on the BM when the waves travel from the staple to the apex, with takes about 10ms for the whole distance.

 \medskip

If the impulse at the maximum amplitude of the wave form undergoes a multiple synchronization to adjust time and phase delays to a synchronized impulse or nervous burst, then further studies would need to consider the time series of the sound and the impulse pattern of the entering wave form further. Such an impulse pattern is regular within a steady-state sound, but may be very complex within initial transients. As musical instruments are mainly identified via their initial transients, such an impulse pattern representation of the sound would be useful, as it would be transmitted to the auditory system as a burst pattern of neurons representing the sound.

\newpage


\begin{thebibliography}{99}

\bibitem{Allen1977} J.B. Allen, ``Two-dimensional cochlear fluid model: New results,'' \emph{J. Aocust. Soc. Am.} \textbf{61}, 110-119 (1977).

\bibitem{Babbs2011} Ch.F. Babbs, ``Quantitative Reappraisal of the Helmholtz-Guyton Resonance Theory of Frequency Tuning in the Cochlea,'' \emph{J. of Biophysics} \textbf{ID 435135} (2011), pp. 1--16.

\bibitem{Bader2015} R. Bader, ``Phase synchronization in the cochlea at transition from mechanical waves to electrical spikes,´´ \emph{Chaos} \textbf{25}, 103124 (2015).

\bibitem{Bader2013} R. Bader, \emph{Nonlinearities and Synchronization in Musical Acoustics and Music Psychology} (Springer-Verlag, Berlin, Heidelberg, Current Research in Systematic Musicology, vol. 2,  2013) pp. 157--284.

\bibitem{Bader2010} R. Bader, ``Buddhism, Animism, and Entertainment in Cambodian Melismatic Chanting smot,´´ In: A. Schneider, A. von Ruschkowski (eds.): \emph{Hamburg Yearbook of Musicology} \textbf{28}, 283-305 (2011).

\bibitem{Bader2005} R. Bader, \emph{Computational Mechanics of the Classical Guitar} (Springer-Verlag, Berlin, Heidelberg, 2005), pp. 51--72.

\bibitem{deBoer1991} E. de Boer, ``Auditory physics. Physical principles in hearing theory,'' \emph{Phys. Rep.}, \textbf{203}, 127-229 (1991).


\bibitem{Brown2000} R. Brown and L. Kocarev, ``A unifying definition of synchronization for dynamical systems,'' \emph{Chaos. Interdiscipl. J. Nonlinear Sci.} \textbf{10(2)}, 344–349 (2000).

\bibitem{Cariani1999} P. Cariani, ``Temporal Coding of Periodicity Pitch in the Auditory System: An Overview,'' \emph{Neural Plascitity} \textbf{6 (4)}, 142 -172 (1999).

\bibitem{Cariani2001} P. Cariani, ``Temporal Codes, Timing Nets, and Music Perception,'' \emph{J. New Music Research} \textbf{30 (2)}, 107–135  (2001).


\bibitem{Goldstein 1973} Goldstein, J.L.: An
Optimum Processor Theory for the Central Formation of the Pitch of
Compex Tones. \emph{J. Acoust. Soc. Am.} 54, 1496-1516, 1973.

\bibitem{Goldstein1978} Goldstein,
J.L., Gersen, A., Srulovicz, P., \& Furst, M.: Verification of the
Optimal Probabilistic Basis of Aural Processing of Pitch of Complex
Tones. \emph{J. Acoust. Soc. Am.} 63, 486-497, 1978.

\bibitem{Helmholtz1863} H. von Helmholtz, \emph{Die Lehre von den Tonempfindungen als physiologische Grundlage f\"ur die Theorie der Musik [On the sensation of tone],} Vieweg, Braunschweig (1863), pp. 1--600.

\bibitem{Hubbard1996} A.E. Hubbard and D.C. Mountain, ``Analysis and Synthesis of Cochlear Mechanical Function Using Models,'' in \emph{Auditory Computation}, L. H. Hawkins, T. A. McMullen, A. N. Popper and R. R. Fay, Editors, Springer Handook of Auditory Research, Springer, New York (1996), pp.  62--120.

\bibitem{Joris1994}
P.X. Joris, L.H. Carney, P.H. Smith and T.C.T. Yin, ``Enhancement of neural synchronization in the anteroventral cochlear nucleus. I. Responses to tones at the characteristic frequency,'' \emph{J. Neurophysiol.} \textbf{71 (3)}, 1022--1036 (1994).

\bibitem{Keidel1975} W. D. Keidel (ed.), \emph{Physiologie des Gehörs : akustische Informationsverarbeitung [Physiology of the ear : acoustical information processing],} Thieme, Stuttgart (1975), pp. 1--409.

\bibitem{Kanis1996} L.J. Kanis and E. de Boer, ``Comparing frequency-domain with time-domain solutions for a locally active nonlinear model of the cochlea,'' \emph{J. Acoust. Soc. Am.} \textbf{100 (4)}, 2543-2546 (1996).

\bibitem{Klapuri2006} A. Klapuri, \emph{Signal processing methods for music transcription,} Springer, New York (2006), pp. 1-440.

\bibitem{Kolston1996} P. J. Kolston and J. F. Ashmor, ``Finite element micromechanical modeling of the cochlea in three dimensions,'' \emph{J. Acoust. Soc. Am.} \textbf{99 (1)} 455-467, (1996).

\bibitem{Licklider1956} J.C.R. Licklider, ``Auditory frequency analysis,'' in \emph{Information Theory}, C. Cherry, Editor, (Butterworth, London, 1956), pp. 253--268.

\bibitem{Louage2004} D.H. Louage, M. van der Heijden and Ph.X. Joris, ``Fibers in the trapezoid body show enhanced synchronization to broadband noise when compared to auditory nerve fibers,'' in \emph{Auditory Signal Processing: Physiology, Psychoacoustics, and Models}, D. Pressnitzer, A. de Cheveigne, St. McAdams and L. Collet, Editors, Springer, New York (2004), pp. 100--106.

\bibitem{Lyon1996}
R. Lyon and S. Shamma, ´´Auditory representation of timbre and pitch,'' in \emph{Auditory Computation}, H. L. Hawkins, T. A. McMullen, A. N. Popper and R. R. Fay, Auditory representation of timbre and pitch, Editors, Springer Handbook of Auditory Research, New York (1996), pp. 221--270.

\bibitem{Meddis1991} R. Meddis and M. J. Hewitt, ``Virtual pitch and phase sensitivity of a computer model of the auditory periphery. II:Phase sensitivity,´´ \emph{J. Acoust. Soc. Am.} 89 (6), 2883-2894 (1991).

\bibitem{Mammano1992} F. Mammano, R. Nobili, ``Biophysics of the cochlea: Linear approximation,'' \emph{J. Acoust. Soc. Am.} \textbf{93 (6)}, 3320-3332 (1992).

\bibitem{Manoussaki2008} D. Manoussaki, R.S. Chadwick, D.R. Ketten, J. Arruda, E.K. Dimitriadis and J.T. O’Malley, ``The influence of cochlear shape on low-frequency hearing,'' \emph{PNAS} \textbf{105 (16)}, 6162-6166 (2008).

\bibitem{Neely1981} S. T. Neely, ``Finite-Difference solution of a 2-dimensional model of the cochlea,'' \emph{J. Acoust. Soc. Am.} \textbf{69 (5)}, 1363-1393 (1981).

\bibitem{Nobili1996} R. Nobili and F. Mammanou, ``Biophysics of the cochlea II: Stationary Nonlinear phenology,'' \emph{J. Acoust. Soc. Am.} \textbf{99 (4)}, 2244-2255 (1996).

\bibitem{Moore2001} D. H. Moore, J. W. H. Schnupp, A. J. King: ``Coding the temporal structure of sounds in auditory cortex,´´ \emph{Nature Neuroscience} \textbf{4} (11), 1055--56 (2001).

\bibitem{Parthasarathi2000} A.A. Parthasarathi, K. Grosh and A.L. Nuttall, ``Three-dimensional numerical modeling for global cochlear dynamics,'' \emph{J. Acoust. Soc. Am.} \textbf{107 (1)}, 474-485 (2000).

\bibitem{Patterson1995} R.D. Patterson, M.H. Allerhand and Ch. Gigu\`ere, ``Time-domain modeling of peripheral auditory processing: A modular architecture and a software platform,'' \emph{J. Acoust. Soc. Am.} \textbf{98 (4)}, 1890--1894 (1995).

\bibitem{Ramamoorthy2010} S. Ramamoorthy, D.-J. Zha and A.L. Nuttall, ``The Biophysical Origin of Traveling-Wave Dispersion in the Cochlea,'' \emph{Biophysical Journal} \textbf{99}, 1687–1695 (2010).

\bibitem{Riemann1880} H. Riemann, \emph{Handbuch der Harmonielehre [Handbook of harmonics],} Leipzig 1880, pp. 1-234.

\bibitem{Schouten1962} J.F. Schouten, R. J. Ritsma and B.L. Cardozo, ``Pitch of the residue,'' \emph{J. Acoust. Soc. Am.} \textbf{34}, 1418-1424 (1962).

\bibitem{Steele1979} C.R. Steele and L.A. Taber, ``Comparison of WKB and finite difference calculations for a two-dimensional cochlear model,'' \emph{J. Acoust. Soc. Am.} \textbf{65}, 1001-1006 (1979).

\bibitem{Szalai2013} R. Szalai, A. Champneys, M. Homer, D.\'O Maoil\'eidigh, H. Kennedy and N. Cooper, ``Comparison of nonlinear mammalian cochlear-partition models,'' \emph{J. Acoust. Soc. Am.} \textbf{133 (1)}, 323-336 (2013).

\bibitem{Terhardt1979} E. Terhardt, ``Calculating virtual pitch,´´ \emph{Hearing Research} \textbf{1}, 155--182 (1979).

\bibitem{Verhulst2010} S. Verhulst, T. Dau and Ch.A. Shera, ``Nonlinear time-domain cochlear model for transient stimulation and human otoacoustic emission,'' \emph{J. Acoust. Soc. Am.} \textbf{132 (6)}, 3842-3846 (2012).

\end{thebibliography}
  \end{document}